\documentclass[prb,reprint,superscriptaddress]{revtex4-1} 
\usepackage{amsmath}  
\usepackage{graphicx} 
\usepackage{amssymb} 
\usepackage{amsfonts} %
\usepackage{graphicx} 
\usepackage{hyperref}

\begin{document}

\title{Macroscopic Brownian Motion on a Chaotic Fluid Interface}


\author{Jack-William Barotta}
\email{jack-william\_barotta@brown.edu} 
\affiliation{
 School of Engineering, Center for Fluid Mechanics,
 Brown University, 184 Hope Street, Providence, RI 02912}

\author{Caroline M. Barotta}
\affiliation{Department of Physics, Brandeis University, Waltham, MA 02453}

\author{Eli Silver}
\affiliation{
School of Engineering, Center for Fluid Mechanics,
Brown University, 184 Hope Street, Providence, RI 02912}

\author{Daniel M. Harris}
\affiliation{
School of Engineering, Center for Fluid Mechanics,
Brown University, 184 Hope Street, Providence, RI 02912}

\date{\today}

\begin{abstract} 
Brownian motion is the erratic motion of an object due to collisions with the fluid in which it is immersed. In this work, we detail a tabletop laboratory demonstration of underdamped Brownian motion wherein a macroscopic particle resting on a driven fluid interface exhibits ballistic motion at short times and diffusive motion at long times. We observe the trajectory of a millimetric disk driven by a field of chaotic Faraday waves excited by a shaker. The crossover from ballistic to diffusive motion occurs at time and length scales experimentally accessible through particle tracking of a video recorded with a standard phone camera. Along with representative data, we provide a complete assembly guide, and operating procedure for students so that the experiment can be readily applied in the classroom. The tabletop setup can also be adapted for other student projects and active research topics relating to particle motion on a vibrating fluid interface. 
\end{abstract}

\maketitle

\section{Introduction}

Robert Brown first observed the random motion of microscopic particles in 1827 when watching pollen grains dance under a microscope. The pollen grains moved about erratically, appearing to be pushed and pulled by an ``invisible" force.\cite{brown1828xxvii} Although Brown ultimately deduced that this effect was not biological in origin by repeating his experiment with inert objects like dust, glass, and soot, a theory for the motion was not developed until Albert Einstein's seminal work in 1905.\cite{einstein1905molekularkinetischen, einstein1956investigations} Einstein deduced that the irregular particle motion documented almost one hundred years prior was in fact due to the collisions of the grain with molecules of water. In addition to moving irregularly, he also suggested that the particles should move {\it ballistically} between the collisions with water molecules that reorient its motion. Under Brown's experimental conditions, however, such ballistic motion would occur over an extremely short timescale, unobservable with the microscopy techniques of the time. As such, the particle would simply appear to diffuse with a diffusion coefficient $D$ that can be related to the mean-squared particle displacement (in two dimensions) as
\begin{equation}
\label{eq: 1d}
    \langle |\Delta\mathbf{r}(t)|^2\rangle = 4Dt.
\end{equation}
A formulation incorporating the effects of particle inertia was derived shortly after Einstein's result by Paul Langevin in 1908, now referred to as the Langevin equation.\cite{langevin1908theorie} Finite particle inertia introduces a crossover timescale $\tau_c$, which delineates the transition from ballistic ($|\Delta\mathbf{r}(t)|^2\sim t^2$) to diffusive motion ($|\Delta\mathbf{r}(t)|^2\sim t$, i.e. Eq. \ref{eq: 1d}). This timescale represents an inertial relaxation period which tends to zero in the fully overdamped (i.e. inertialess or high friction) limit.

To get a sense of scale, it is instructive to compute an estimate for this crossover timescale, $\tau_c= \frac{m}{\gamma}$, where $m$ is the mass of the particle and $\gamma$ is the friction factor.\cite{ramaswamy2000pollen} For a pollen grain, we can estimate the friction factor as $\gamma = 6\pi \mu R$ from Stokes' drag\cite{stokes1851effect}, where $\mu$ is the liquid viscosity and $R$ is the particle radius. Taking the mass of the grain as $m = \frac{4}{3}\pi\rho_sR^3$, we can estimate the timescale as $\tau_c = \frac{ 2 \rho_{s} R^2}{9\mu}$. Assuming that the pollen grain is suspended in water ($\rho = 1$ g/cm$^3$, $\mu = 1$ mPa$\cdot$s), is neutrally buoyant ($\rho_s=\rho$), and has radius $R\approx$ 1 $\mu$m, we can estimate the crossover timescale for Brown's experiments to be $\tau_c \approx 0.1$ $\mu$s, requiring a temporal resolution on the order of nanoseconds to observe. As such, this crossover at the microscopic scale was not experimentally documented until over 100 years after Einstein's prediction.\cite{huang2011direct, li2010measurement}

Analog macroscopic demonstrations of Brownian motion present an opportunity to more easily observe diffusive dynamics without complex optical setups, readily implementable in a minimal physics laboratory. In a recent AJP article, Brigante \textit{et al}. document a tabletop experiment that allows students to probe the assumptions and dynamics of Brownian motion using vibrated granular matter.\cite{brigante2024experimentation}  The authors use a large bead ($R=3$ mm) immersed in a background of small beads ($R=0.45$ mm), which play the role of the pollen grain and water molecules, respectively, of Brown's original experiment. The collection of beads is vertically vibrated on a loudspeaker, with diffusive behavior of the large particle emerging as a result of the chaotic collisions with the surrounding smaller particles. While their highly frictional granular system allowed for careful probing of the diffusive regime, it was not demonstrated to access the short-time ballistic (inertial) dynamics.

To complement that work, we here present an alternate version of the experiment, replacing the small beads with a periodically driven water bath, that allows us to elucidate the short-time ballistic dynamics of Brownian motion. The use of the water environment allows us to reduce the particle friction to a level where the inertial dynamics become readily observable.
When a liquid interface is vertically oscillated above a critical amplitude, nonlinear subharmonic Faraday waves form on the free surface. An unsteady wave field and surface flows result from the driving. When a floating object is placed on the bath, the waves and associated surface flows randomly move the object around in a way that can be approximated as a white noise random forcing under particular experimental conditions. This system has been well-documented in recent research as a simple platform that behaves as a macroscopic 2D ideal gas\cite{welch2014ballistic}, and also exhibits 2D turbulence.\cite{xia2019tunable, yang2019passive, francois2018rectification, yang2019diffusion} Importantly, the particle's non-negligible inertia increases the crossover time, allowing the experimentalist to readily probe ballistics dynamics without requiring specialized equipment. This experiment highlights the balance between inertia and drag in setting the system's finite memory, captured via the ballistic timescale. 

Fluid-shaker experimental setups have been central to a number of research topics in the last decade, with investigations demonstrating the rich interplay between particle dynamics and surface waves and flows.\cite{welch2014ballistic, xia2019tunable, yang2019passive, francois2018rectification, yang2019diffusion, thomson2023nonequilibrium} Transcending fundamental fluid mechanics, objects on vibrating liquid baths have been also shown to demonstrate a large assortment of emergent behaviors, with self-propelled droplets acting as hydrodynamic quantum analogs\cite{couder2005walking, bush2024perspectives}, and asymmetric solid floaters (referred to as ``capillary surfers'') providing a tunable system to explore topics and open questions in active matter.\cite{harris2025propulsion} 

In this note, we describe the application of the vibrating liquid interface to the underdamped dynamics of a randomly forced object on the interface. Students can observe the motion of an inertial particle as it is randomly pushed about by supercritical Faraday waves, and directly quantify the ballistic and diffusive motion of the object on timescales accessible with a standard camera. Pedagogically, this setup enables students to make a direct physical connection between the physics of ballistic motion and diffusive motion, with the latter often only presented abstractly as a phenomenon exclusive to the microscale. The experiment can be introduced in conjunction with the Langevin equation to relate the statistics of the floater's motion directly to theoretical predictions. We provide all experimental design and software files necessary to replicate the experiment and analysis. The experiment we describe also serves as an accessible, hands-on introduction to particle motion on a chaotic fluid interface, naturally forming a link between fundamental concepts in standard physics curricula and active areas of ongoing research.

\section{Experimental Setup}

\begin{figure}
    \centering
    \includegraphics{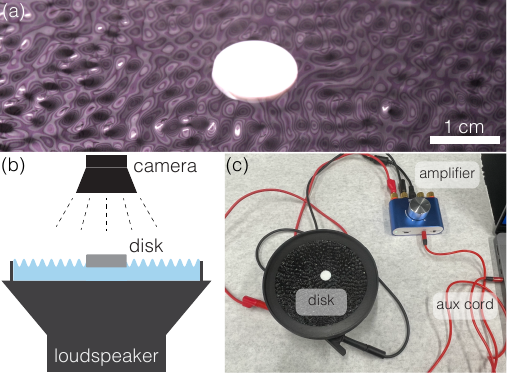}
    \caption{(a) Photograph of a floating disk of radius $R=6$ mm on an air-water interface of Faraday waves forced at $f=180$ Hz, visualized with a reflective color pattern.\cite{harris2017visualization} (b) Schematic of the experimental setup. A 3D-printed bath filled with water is hot glued to a speaker. As the amplitude of speaker oscillations is increased by increasing the volume, waves form on the surface that mimic a random 2D forcing, moving the disk around the bath. (c) Photograph of the experiment setup.}
    \label{fig:setup}
\end{figure}

The setup consists of a speaker, water bath, amplifier, and an object that floats on the surface of the water-filled bath (Fig. \ref{fig:setup}). While a variety of different materials, methods, and sizes can be used for the object, we use white 3D-printed disks  ($R=6$ mm). The bath ($R_{\text{b}} = 5$ cm, uniform depth $H=0.5$ cm) is 3D-printed using black filament to provide suitable contrast with the white floating disk and thus enable robust particle tracking. The floating disk rests partially submerged on the air-water interface via a balance of buoyancy, weight, and surface tension. The 3D-print source files, assembly and setup instructions, and operating procedures can be found in the associated Github.\cite{github}

\begin{figure}
    \centering
    \includegraphics{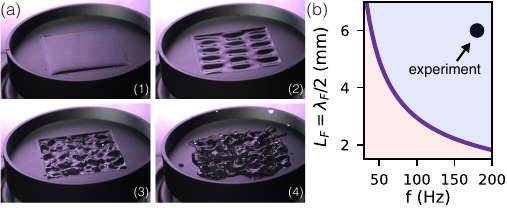}
    \caption{(a) The four qualitative regimes of Faraday wave behavior demonstrated in a small square bath with sides of 4 cm at $f=60$ Hz: (1) still, (2) standing, (3) erratic, and (4) drop emission. (b) The forcing lengthscale as a function of frequency for water in a dish of depth $H= 0.5$ cm. We choose disks that are larger than the forcing lengthscale, i.e. $R > L_F \equiv \lambda_F /2$. The experimental conditions presented herein are indicated by the black dot.}
    \label{fig:faraday}
\end{figure}

In order to generate motion on the interface, students generate a single frequency tone with a laptop or cell phone (we use the website \href{https://onlinetonegenerator.com/}{onlinetonegenerator.com}). If available, a standalone benchtop signal generator can also be used. Above a critical oscillation amplitude, controlled simply by adjusting the output volume, the fluid surface destabilizes into nonlinear surface waves referred to as Faraday waves, first documented by Michael Faraday in 1831.\cite{faraday1831xvii} As the amplitude is increased for a fixed frequency, the surface will qualitatively evolve from 1) a flat, unperturbed surface to 2) standing orderly Faraday waves to 3) erratic unsteady Faraday waves that move and scar on the surface to 4) a surface that begins splashing and emitting droplets (Fig. \ref{fig:faraday}(a)).

In the erratic regime (regime 3), the Faraday wave crests and troughs appear to move chaotically around the bath as a function of time. In response, the particle on the surface randomly moves around the bath, with the Faraday waves and flows acting as small ``particles'' colliding with our large floating object. The dominant wavelength of the waves (which also dictates the characteristic lengthscale of the surface flows) represents a tunable lengthscale that we can set via the frequency of the tone. As noted in Brigante \textit{et al}.\cite{brigante2024experimentation}, the ``particles'' contributing the random forces should be small relative to the object size for effective analogy with Brown's original experiments. In their study, the large bead was approximately $6\times$ the size of the small bead. For our experiment, we ensure that the size of the object $R$ is greater than the size of the fluctuation lengthscale\cite{xia2019tunable} $L_F = \lambda_F/2$, where $\lambda_F$ is the characteristic wavelength. In this regime, a relationship reminiscent of the fluctuation-dissipation relation is known to hold.\cite{xia2019tunable} For Faraday waves, the wavenumber $k_F = 2\pi/\lambda_F$ is related to the driving frequency $f$ via
    \begin{equation}
    \label{eq: wavenumber}
        \left(\pi f\right)^2 = \left( gk_F + \frac{\sigma}{\rho}k_F^3\right)\tanh(k_FH)
    \end{equation}
where $\sigma$ is the surface tension and $g$ is the acceleration due to gravity. In the limit that $k_FH \gg 1$, $\tanh(k_FH) \approx 1$, allowing for a polynomial solver to be used to compute the wavelength\cite{desmos} as the sole real solution to Eq. \ref{eq: wavenumber}. The forcing lengthscale as a function of frequency is shown in Fig \ref{fig:faraday}(b), with admissible disk sizes falling within the blue shaded region.

We use a standard phone camera acquiring video at 30 fps to record 12 independent video clips of the disk motion on the bath viewed from above. We use a frequency of $f=180$ Hz with a 3D-printed disk of $R=6$ mm and thickness $d=2$ mm. For these parameters, we anticipate that the ballistic-to-diffusive transition occurs on the order of $0.1$ s based on prior studies with similar experimental conditions.\cite{welch2014ballistic, xia2019tunable} We record videos for 30 seconds, initially placing the disk near the center of the bath, in order to capture that transition while also avoiding complications arising from confinement by the bath, which causes the observed mean squared displacement (MSD) to plateau over time.\cite{welch2014ballistic, brigante2024experimentation} Alternatively, other authors have omitted trajectories which were within some specified distance of the walls of the container to avoid confinement effects. We track the particle's position in time from the recordings with a code written in Python that uses the $\mathtt{OpenCV}$ package (see Github\cite{github} for code and examples). Alternatively, students could use Tracker, another open-source software for particle tracking.\cite{brown2008video} By tracking the planar position ($\mathbf{x}=\left\{x,y\right\}$) of the particle as a function of time, we can compute the velocity, $\mathbf{v} = \Delta \mathbf{x} / \Delta t$, where $\Delta t$ is the inverse of the filming frame rate (in frames per second). 

\begin{figure}
    \centering
    \includegraphics{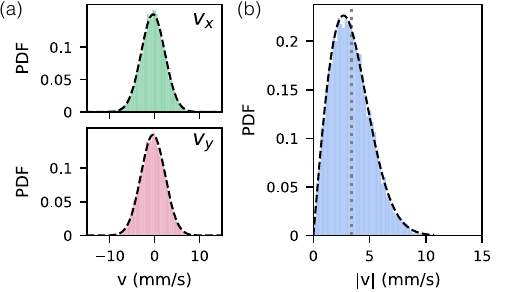}
    \caption{(a) The PDF of the $x$ and $y$ components of the velocity over 12 independent trials, with the best-fit Gaussian (black dashed line). The speed of the particle $|\mathbf{v}| = \sqrt{v_x^2+v_y^2}$ follows a Rayleigh distribution (black dashed line). The mean is marked with the gray dashed line.}
    \label{fig:vels}
\end{figure}

The probability density function (PDF) of the $x$ and $y$ components of the object's velocity are Gaussian (Fig. \ref{fig:vels}(a)), and as such the speed of the particle follows a Rayleigh distribution (Fig. \ref{fig:vels}(b)). The particle's mean speed is $3.4$ mm/s, indicating that the particle is moving approximately $0.5$ radii per second. 

We also compute the velocity auto-correlation function (VACF) and mean-squared displacement (MSD). The VACF measures the correlation between the object's velocity at one instant in time $\mathbf{v}(t)$ with the object's velocity at a later time ($\mathbf{v}(t+\Delta t)$). Mathematically, the VACF over a single trial can be expressed as
\begin{equation}
\label{eq: VACF_exp}
    \text{VACF}(\Delta t) = \langle \mathbf{v}(t) \cdot \mathbf{v}(t+\Delta t) \rangle,
\end{equation}
where the use of $\langle \cdot \rangle$ indicates the averaging over multiple realizations in time.  At $\Delta t=0$, the VACF is simply the mean speed squared $\langle |\mathbf{v} |^2\rangle$.

The MSD measures how far a particle moves from its original position over a duration of time $\Delta t$. We compute the MSD for our experimental data as
\begin{equation}
\label{eq: MSD_exp}
    \text{MSD}(\Delta t)  = \langle |\mathbf{r}(t+\Delta t) - \mathbf{r}(t) |^2 \rangle.
\end{equation}
The larger the MSD, the farther the particle has moved from its original position within time $\Delta t$. We compute both the VACF and MSD for 12 independent trials and average over all trials (Fig. \ref{fig: stats}). We relate the experimental results to corresponding theoretical predictions derived from the Langevin equation, discussed next.

\section{Modeling the Motion}

The object's motion in the plane can be modeled via Newton's second law to arrive at the Langevin equation
\begin{equation}
    m\dot{\mathbf{v}} + \gamma \mathbf{v} = \mathbf{F}(t),\label{eqn:langevin}
\end{equation}
where $m$ is the mass, $\gamma$ is the friction factor, and $\mathbf{F}(t)$ is a random forcing modeling the Faraday waves and concomitant surface flows. We assume for simplicity that the object experiences a drag force that is linear in velocity, although other forms of hydrodynamic drag could lead to alternative scalings with velocity.

The strength of the random forcing is dependent on the effective ``temperature'' of the system, which here is not thermal temperature, but rather is related to the amplitude of the wavefield (see Welch {\it et al.} \cite{welch2014ballistic} for an extended discussion). We can rewrite the Langevin equation (Eq. \ref{eqn:langevin}) in terms of the crossover time $\tau_c = \frac{m}{\gamma}$ and the diffusion coefficient $D$ to get
\begin{equation}
\label{eq:LE}
    \tau_c\dot{\mathbf{v}} + \mathbf{v} = \sqrt{2D}\mbox{\boldmath$\eta$}(t),
\end{equation}
where $\mbox{\boldmath$\mbox{\boldmath$\eta$}$}(t)$ is white noise with the properties
\begin{equation}
   \left\langle \mbox{\boldmath$\eta$}(t)\right\rangle =\mathbf{0} \quad \quad  \langle \eta_i(t) \eta_j(t') \rangle =
\begin{cases} 
\delta(t-t') & \text{if } i=j \\[2mm]
0 & \text{if } i \neq j
\end{cases},
\end{equation}
where $\eta_i(t)$ is the $i^{\text{th}}$ component of the noise vector, and $\delta(t-t')$ is the Dirac delta function. We can express both the VACF and MSD in terms of $\tau_c$ and $D$ via
\begin{equation}
\label{eq: VACF}
\text{VACF}(\Delta t) = \langle |\mathbf{v} |^2\rangle \text{e}^{-\Delta t/\tau_c} = \frac{2D}{\tau_c}\text{e}^{- \Delta t/\tau_c},
\end{equation}
and
\begin{equation}
\label{eq: MSD}
    \text{MSD}(\Delta t) = 4D\tau_c\left( \frac{\Delta t}{\tau_c} - 1 + e^{-\frac{\Delta t}{\tau_c}}\right).
\end{equation}
Eq. \ref{eq: MSD} can be expanded in both the short-time $(t \ll \tau_c)$ and long-time $(t \gg \tau_c)$ limits, yielding asymptotic predictions for the ballistic and diffusive regimes respectively:
\begin{equation}
\label{eq:approxs}
MSD(\Delta t) \approx 
\begin{cases}
\frac{2D}{
\tau_c} (\Delta t)^2 & \Delta t \ll \tau_c, \\[1.2em]
4 D (\Delta t) & \Delta t \gg \tau_c.
\end{cases}
\end{equation}

\begin{figure}
    \centering
    \includegraphics{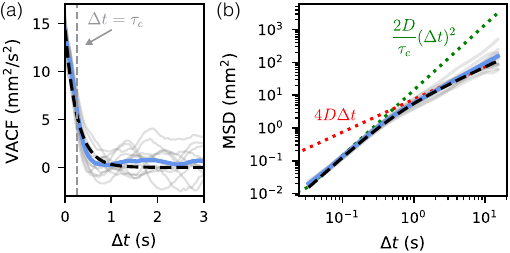}
    \caption{(a) The velocity auto-correlation function (VACF) as a function of time. The fitted value of $\tau_c$ (gray dashed line) captures the exponential decay of the VACF. (b) The mean-squared displacement (MSD) as a function of the time. In both cases, independent trials are shown in solid gray and the mean across the 12 trials is shown in blue. The best-fit curve (black dashed line) for both the VACF (Eq. \ref{eq: VACF}) and MSD (Eq. \ref{eq: MSD}) using the fitted value for the crossover time $\tau_c$ and resultant diffusion constant $D$ is overlaid. In (b), the limiting cases for ballistic (green) and diffusive (red) motion are shown from Eq. \ref{eq:approxs}.}
    \label{fig: stats}
\end{figure}

There are numerous ways to use the experimental VACF and/or MSD to extract numerical values for both $\tau_c$ and $D$. We fit the mean VACF averaged across all trials to the corresponding theoretical expression (Eq. \ref{eq: VACF}, Fig. \ref{fig: stats}(a)). Using $\text{VACF}(\Delta t) = \langle |\mathbf{v}|^2 \rangle e^{-\Delta t/\tau_c}$, we first fit the crossover time $\tau_c$. With the fitted value of $\tau_c$, we then rearrange the two expressions for VACF$(0)$ in Eq. \ref{eq: VACF_exp}, obtaining $D = \langle |\mathbf{v}|^2 \rangle \tau_c/2$. Through this procedure. we find that $\tau_c = 0.26\pm 0.034$ s and $D = 1.9 \pm 0.24$ mm$^2$/s. We then use these values determined from the velocity statistics to predict the MSD evolution via Eq. \ref{eq: MSD}, showing excellent agreement with the experiment (Fig. \ref{fig: stats}(b)). While we chose to fit the parameter $\tau_c$ via VACF and then demonstrate consistency with the MSD, alternative fitting methods could use the short- and long-time approximations for the MSD to obtain values for $\tau_c$ and $D$ using Eq. \ref{eq:approxs}. On a log-log plot, visual inspection of the MSD identifies a crossover in the slope roughly near $0.1-0.5$ s, where the slope (power-law exponent) transitions from ballistic (slope of 2) to diffusive (slope of 1), consistent with our best-fit value $\tau_c=0.26$ s. We expect that for differing values of disk radius, driving amplitude (volume), and driving frequency, the values of $\tau_c$ will remain in the range of $ 0.1 \lesssim \tau_c \lesssim 1$ s, and the value of $D$ within $1 \lesssim D \lesssim 20$ mm$^2$/s. 

\section{Conclusion}

We have presented a macroscopic experiment for introducing and exploring underdamped Brownian motion in a tabletop setup, and in particular, one that effectively highlights the transition between ballistic and diffusive motion for inertial particles.\cite{welch2014ballistic} Our accessible system uses Faraday waves and their associated surface flows to erratically force a floating particle along the fluid surface, leading to observable ballistic and diffusive motion that can be captured with a standard phone camera and analyzed with open-source particle tracking software. Supplemental documents outline the experimental design and setup with associated source files, as well as provide corresponding code for video processing and parameter estimation, all of which can be readily adapted to curricula introducing Brownian motion.

The setup is naturally tunable given the design choices of driving parameters (frequency, amplitude/volume), fluid properties, and floater geometry, easily bridging into further studies. The experiment can be extended to consider asymmetric particles, calculating their rotational diffusion or observing their self-propulsion which emerges via the rich interplay between particle and wave coupling.\cite{yang2019passive} Further, one could easily move to a regime where the object is smaller than the forcing lengthscale by changing the object size and/or driving frequency. In this regime, the fluctuation-dissipation-like relation breaks down, and the object becomes strongly embedded into the surface flow itself, behaving more similar to a fluid tracer.\cite{xia2019tunable} Finally, our setup could also be adapted to study the influence of confinement or particle-particle interactions, and to applications related to current research topics on particle motion and aggregation in active baths.\cite{thomson2023nonequilibrium, ray2023rectified}

\section*{Author Declaration}
The authors have no conflicts to disclose.

\bibliographystyle{apsrev}
\bibliography{refs.bib}

\end{document}